\documentclass[prc,showpacs,onecolumn,floatfix]{revtex4}
\usepackage{graphicx}
\usepackage{dcolumn}
\usepackage{bm}
\usepackage{multirow}
\usepackage{color}
\usepackage{amssymb}
\usepackage{natbib}
\begin{document}
\title{Antikaons in neutron star studied with recent versions of relativistic mean-field models}

 \author{Neha Gupta and P.Arumugam}
 \address{Department of Physics, Indian Institute of Technology Roorkee, Roorkee - 247667 India}
 
\begin{abstract}
We study the impact of additional couplings in the relativistic mean field (RMF) models, in conjunction with antikaon condensation, on various neutron star  properties. We analyze different properties such as in-medium antikaon and nucleon effective masses, antikaon energies, chemical potentials and the
mass-radius relations of neutron star (NS). We calculate the NS properties
with the RMF (NL3), E-RMF (G1, G2) and FSU2.1 models, which are quite successful
in explaining several finite nuclear properties. Our results show that the onset of kaon condensation in NS strongly depends on the parameters of the Lagrangian, especially the additional couplings which play a significant role at higher densities where antikaons dominate the behavior of equation of state. 
\end{abstract}
\maketitle

\section{Introduction}
This work aims at furthering our understanding of the interior of neutron stars (NS). NS star  as the name implies, is mostly composed of neutrons.  However, with increasing density (i.e. from crust to core of the NS) other exotic particles like hyperons, kaons, pions and quarks can exist by strangeness changing processes \cite{lattimer:426,gle}. In the recent years, the study of NS properties has gained momentum due to several new observations \cite{expt_m,demorest},  and the development of various theoretical models \cite{FSU_para,ermf,furnstahl,neha}.

In NS, antikaon condensation is one of the several possible
transitions that could exist at high density. As the density of the
NS increases,  the effective mass of an in-medium
antikaon will decrease due to the attractive interaction between
antikaons and nuclear matter and, hence at sufficiently high density the antikaons are favoured to condense \cite{Banik035802,Pal553}. This was first demonstrated by Kaplan and Nelson \cite{Kap} within a chiral $SU(3)_L\times
SU(3)_R$ model where baryons directly couple with~antikaons. Later, antikaon  condensation in the core of neutron stars was studied by other groups also \cite{Ellis199511,Brown:355,Thorsson1994693,banik:055805,glen_kaon} and most
of these works suggest a strong influence of antikaon condensation in the
NS properties. We investigate such a condensation from nucleonic phase to antikaonic phase, using recent versions of relativistic mean field (RMF) models which are proved to be successful
in explaining several properties of finite nuclei and neutron stars. In the
presence of such a phase transition we study the effect of additional couplings, in extended RMF models, on the equation of state (EoS) and hence on the NS properties.

We calculate the NS properties with the RMF (NL3 parameter set), the E-RMF
(G1,G2 parameter sets, with four additional couplings to RMF), and the FSU2.1  (with three additional couplings to RMF) models. The additional couplings in the E-RMF and FSU2.1 models  represent the  non-linear interactions between scalar and vector mesons as well as tensor couplings.  The free parameters of the model  Lagrangian have been optimized by fitting to the
ground-state properties of selected magic nuclei \cite{FSU_para,furnstahl}.  Extension of these models with the inclusion of antikaons is the central interest of the present work.  In the next section, we describe the extension of nucleonic Lagrangian and discuss the constraints for the antikaon condensation. Our results are discussed in section \ref{sec:results} which is followed by the conclusions drawn from the present work. 

\section{Formalism} \label{sec:model}
The details of RMF, E-RMF and FSU2.1 models and corresponding Lagrangians for nucleon part, are explained in Refs.~\cite{nl3_para,furnstahl,Shen:065808}, respectively. In this work, we extend them with the inclusion of the antikaons
$K^-$ and $\bar{K^0}$. The Lagrangian for the antikaon part reads
\begin{equation}\label{kaon}\mathcal{L}_K=D_\mu^*K^*D^\mu K-m_k^{*2}K^*K,
\end{equation}
 with $K\equiv K^-$ or $\bar{K^{0}}$.
The scalar and vector fields are coupled to antikaons analogous to the minimal coupling scheme \cite{glen_kaon} via
the relations
\begin{eqnarray}
m_K^*&=&m_K-g_{\sigma K}\sigma\label{eq:effmass}~~~ \textrm{and}
\\
D_\mu&=&\partial_\mu+ig_{\omega K}V_\mu+ig_{\rho K}\tau_3\cdot R_\mu,
\end{eqnarray}
where $m_K$ stands for the antikaon's mass ($m_{K}=495$ MeV) and $\sigma$, $V_\mu$, and $R_\mu$ represent scalar,
vector and isovector fields, respectively. In the mean field approximation, these fields
are replaced by their expectation values $\sigma$, $V_0,$ and $R_0$ and 
the coupling constants corresponding to these fields
are represented by $g_{\sigma K}, g_{\omega K}$ and $g_{\rho K}$.

Energy relations for the antikaons ($K^-$,$\bar
K^0$) are 
\begin{equation}\label{ke}\omega_{K^{-},\bar K^0}=m_K-g_{\sigma K}\sigma-g_{\omega K}V_0\mp g_{\rho K}R_0, \label{eq:ke}
\end{equation}
where $\mp$ sign represents the  isospin projection of the antikaons
$K^-$ and $\bar K^0$, respectively. The above expression suggests that antikaon condensation
is significantly influenced by the $\rho$ meson field or vice-versa.

The constraints involving
chemical potentials and baryon densities can be written as 
\begin{eqnarray}
\mu_n&=&\mu_p+\mu_e, \nonumber\\
\mu_e&=&\mu_\mu, \textrm{\ \ and}\nonumber\\
 q&=&\rho_p-\rho_e-\rho_\mu-\rho_{K^-}.
\label{beta_eq}
\end{eqnarray}
Here $q$ represents the total charge of the NS and it is zero according to the charge neutrality condition.

The total  energy density in the presence of antikaons can be written as 
\begin{equation}\label{ed}\epsilon=\epsilon_N+m_K^*(\rho_{K^-}+\rho_{\bar K^0}),
\end{equation}
where $\epsilon_N$ is the energy density of nucleon phase as given in Ref.~\cite{neha}
and $\rho_{K^-}$, $\rho_{\bar K^0}$ are the densities of antikaons and can be written as,
 \begin{eqnarray}
 \rho_{K^-,\bar K^0}=2(\omega_{K^-,\bar K^0}+g_{\omega K}V_0\pm
 g_{\rho K}R_0)K^* K.
 \end{eqnarray}
In the $s$-wave condensation, unlike the energy density, the expression for  pressure remains same  \cite{glen_kaon}.
The conditions for onset of antikaons are,
$\omega_{K^-}=\mu_e$ for $K^-$ and $\omega_{\bar K^0}=0$  for $\bar K^0$. Using these conditions, we can calculate $\sigma$, $V_0$, $R_0$, $k_{fp}$, $k_{fn}$, $k_{fe}$, $k_{f\mu}$,
$\rho_{K^-}$ and $\rho_{\bar{K^0}}$, for any chosen baryon density. After getting
this solution, we can calculate energy density and pressure (EoS) for the
antikaon
phase. FSU2.1 \cite{Shen:065808} has same parameters as in FSUGold  \cite{FSU_para}, but with one extra
term in the expression for pressure. The detailed list of parameters,
for both kaon-meson and nucleon-meson couplings for G1,G2, NL3 and FSU2.1(FSUGold)
  are  given in Ref.~{\cite{neha}.}

\section{Results and discussions}\label{sec:results}
\begin{figure}[ht]\center
\includegraphics[width=0.5\columnwidth]{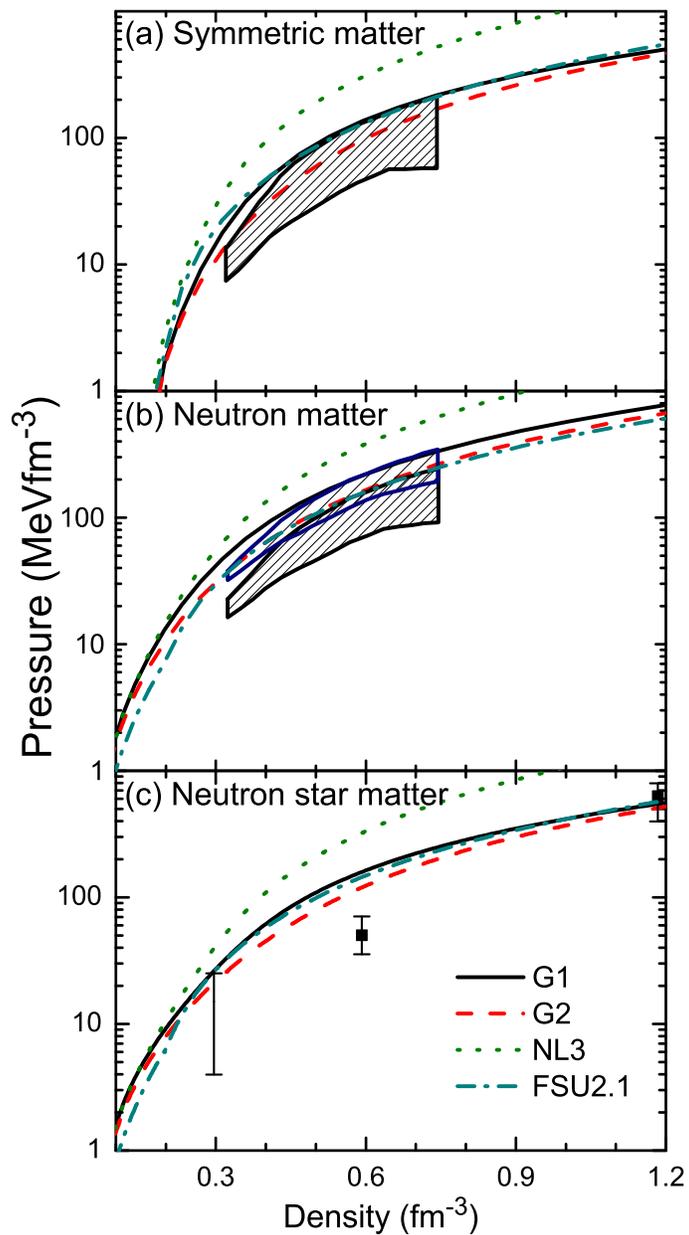}
\caption{Zero temperature EoS for (a) symmetric nuclear matter, (b) pure
neutron
matter and (c) neutron star matter. Shaded area in the first two panels represents the region consistent with the experimental data \cite{expt_data}. Solid squares in the bottom panel represent the
observational extraction \cite{expt_m5}, however, not uniquely constrained
\cite{Fattoyev:055803}.}
\label{fig:denpre1}
\end{figure}
\subsection{Role of additional couplings}
We start with the study of different interactions in explaining symmetric matter, neutron matter and neutron star matter properties. The important observation
 from Fig.~\ref{fig:denpre1} is that the softness in EoS from different
interactions is mainly due to the additional couplings. The additional  couplings dominate at higher densities and yield the EoS consistent
with the flow data \cite{expt_data}. These results improve our confidence
to use the parameters for calculating NS properties where the high density behaviour of the EoS is crucial. Among the different parameter sets
considered here, G2 and FSU2.1 have a positive quartic
scalar self-coupling which is more relevant than a negative one \cite{ermf,kappa_4}.  With increasing neutron fraction, EoS corresponding to FSU2.1 parameter set becomes much softer, which is due to the additional isoscalar-isovector coupling, which suppresses the $\rho$ field. The EoS for NS matter calculated
with G1, G2, and FSU2.1 are consistent with the observational constraints
\cite{expt_m5} which however are not so accurate \cite{Fattoyev:055803}.
\begin{figure}[h]\center
\includegraphics[width=0.8\columnwidth]{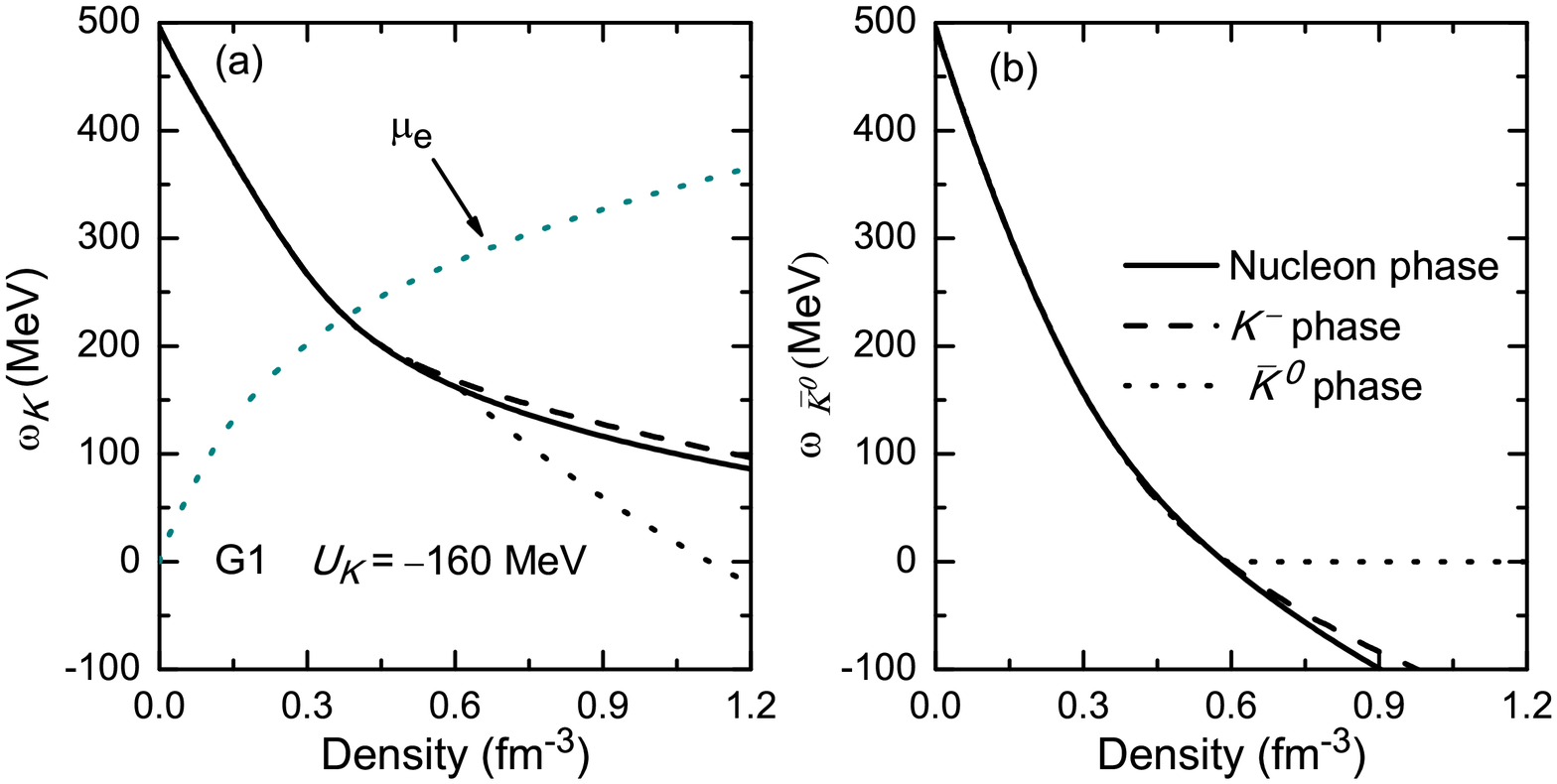}
\caption{The density dependence of (a)$K^-$ and (b)$\bar
K^0$ energies in NS matter calculated with the parameter set G1 and $U_K=-160$ MeV. The nucleon phase comprises the particles $n, p, e^-$ and $\mu^-$, the
$K^-$ phase comprises the particles $n, p, e^-, \mu^-$ and $K^-$ and the $\bar K^0$ phase comprises the particles $n, p, e^-, \mu^-,K^-$ and
$\bar K^0$. The calculations done for nucleon, $K^-$ and $\bar K^0$ phases
are represented by
solid, dash and dot lines, respectively.}
\label{fig:kaonenergy}
\end{figure}
\subsection{Results from G1}
Now we look into the conditions at which the antikaons start to appear in
the  neutron star matter. For this we calculate
the in-medium antikaons' energies  with G1 parameter set, as a function of density (Fig.~\ref{fig:kaonenergy}).  Both the antikaons' energies  decrease with increasing density [Eq.~(\ref{eq:ke})], and the point where $K^-$ energy is less than the electron chemical potential and
the point at which $\bar K^0$ energy is equal to zero represent the onset of $K^-$ and $\bar K^0$ respectively.  The density dependence of $\omega_{K^-}$ is similar to that
of the EoS and electron chemical potential varies in a way similar to the
symmetry energy \cite{neha}. So any change in the density dependence of EoS or that of symmetry energy will affect the
onset as well as the effect of $K^-$ condensation. We observe
similar effects with other parameter sets also.    In general, $\bar K^0$
can appear only at densities higher than the
one at which $K^-$ condenses. This is due to the fact
that, with the onset of $K^-$ condensation, $n\rightarrow p+K^-$ is the most
preferred process, and hence the proton fraction rises dramatically and even
exceeds the neutron fraction at higher densities. With the onset of $\bar
K^0$ condensation, the proton fraction will change completely. There will be
a competition between the processes $N\rightarrow N + \bar K^0$ and
$n\rightarrow p + K^-$, resulting in a perfectly symmetric matter of nucleons and antikaons inside the NS
\cite{Banik035802}. Thus the onset of $\bar K^0$ depends on the onset of $K^-$ which
in turn depends on the value of $U_K$. In our further discussions we restrict to $U_K=-160$ MeV and look into dependencies on other parameters. In the $K^-$ phase [Fig.~\ref{fig:kaonenergy}(a)]
with G1 parameter set,  antikaon energies increase marginally when  compared to nucleon phase.
However in the $\bar K^0$ phase [Fig.~\ref{fig:kaonenergy}(b)], $K^-$ energy decreases and $\bar K^0$ energy
become zero.  

\begin{figure}[t]\center
\includegraphics[width=0.8\columnwidth]{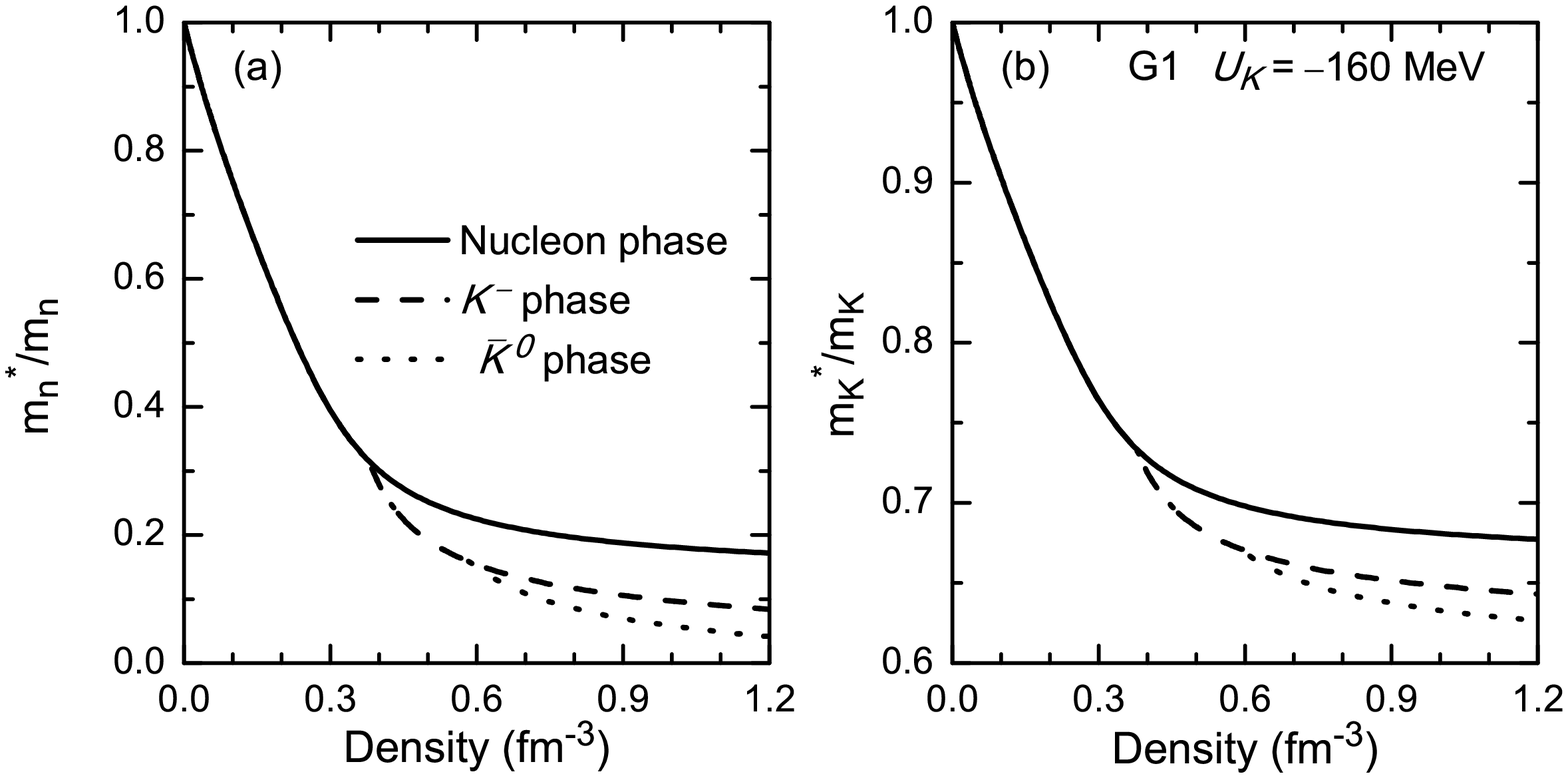}
\caption{The density dependence of 
effective masses of (a) nucleon and (b) antikaon, calculated with different parameter sets ($m_n=939$ MeV, $m_K=495$ MeV). The calculations done for nucleon, $K^-$ and $\bar K^0$ phases
are represented by
solid, dash and dot lines, respectively.}
\label{fig:eff_mass}
\end{figure}

In Figure~\ref{fig:eff_mass}, we present effective-masses of (a) nucleon ($m_n^*/m_n$), and (b) antikaons' ($m_{K}^*/m_{K}$), calculated with G1 parameter set and at
 $U_K=-160$
MeV. As the density of the
NS increases,  the effective mass of an in-medium
antikaon decreases and hence at sufficiently high density, the antikaons can
 condense.  However, the reduction in the effective mass of antikaon is
not strong as compared to that of the nucleon. In the phases of $K^-$ and then $\bar K^0$ these ratios decrease marginally, that is related to scalar field (Eq.~\ref{eq:effmass}) which is not much influenced by the presence of antikaons. We  observe similar results with other parameter sets  considered in this work.  

\subsection{Comparison between different interactions}
\begin{figure}[t]\center
\includegraphics[width=0.5\columnwidth]{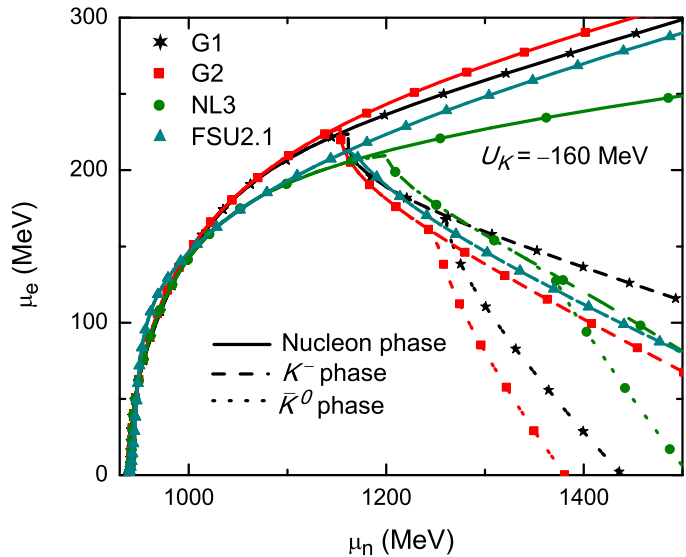}
\caption{The electron and neutron chemical potentials in NS matter calculated with different parameter sets and $U_K=-160$ MeV.}
\label{fig:cp}
\end{figure}

So far we discussed our results with the G1 parameter
set and here we compare results from different interactions considered in this work. In Fig.~\ref{fig:cp}, we show the electron chemical potentials ($\mu_e$) as a function of neutron chemical potentials ($\mu_n$).  $\mu_e$ quantifies the number density of electrons which falls sharply in the presence of antikaons. The first kink in the $\mu_e$ denotes the transition point for nucleon phase to $K^-$ phase and the
second kink represents the transition from $K^-$ phase to $\bar K^0$ phase. The
maximum value of  $\mu_e$ 
is seen to be larger for G1 and G2, and lower for NL3 and FSU2.1 parameter sets.
Both the transitions at which the two antikaons set in, strongly
depend on the interactions. It is clear that for $U_K=-160$ MeV, all the interactions lead only to a
second order phase transition because in Fig.~\ref{fig:cp}, with all the parameter
sets, there is only one value of $\mu_e$
for a given $\mu_n$ and parameter set.

\begin{figure}[t]\center
\includegraphics[width=0.5\columnwidth]{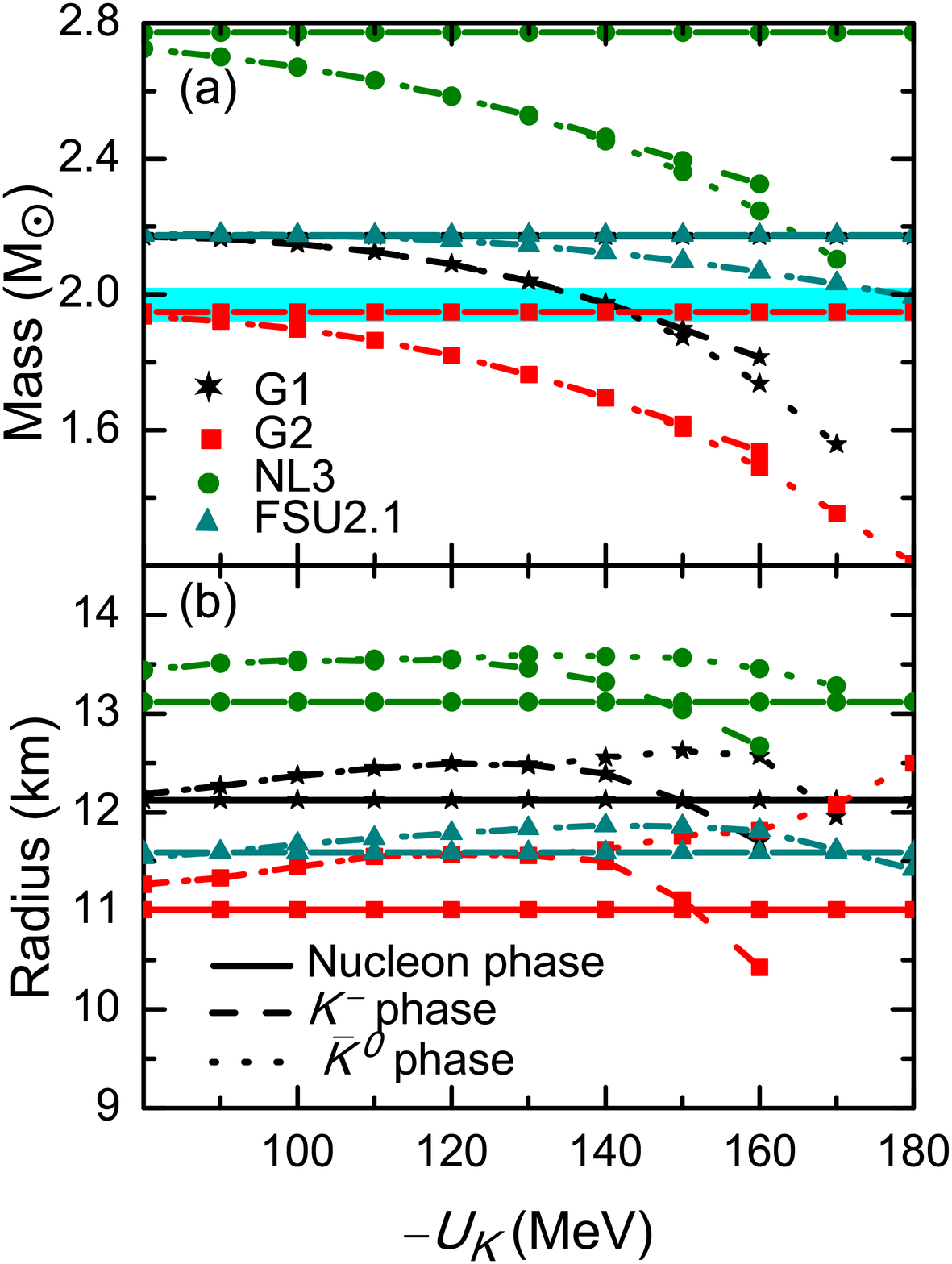}
\caption{The maximum mass and corresponding radius of a NS as a function of kaon
optical potential $U_K$ for nucleon, $K^-$ and $\bar K^0$ phases calculated
with different parameter sets. Mass is given in units of solar mass $M_\odot$. }
\label{fig:massradius}
\end{figure}

The role of antikaons in modifying the EoS is very well reflected in the
results for the mass and radius of NS. With the EoS defined, we can obtain the mass-radius relation by solving the
well-known Tolman-Oppenheimer-Volkoff (TOV) equations
\cite{tov1,tov}. The dependence of maximum mass and corresponding radius, on the kaon optical potential ($U_K$) is depicted
in Fig. \ref{fig:massradius},  where the calculations are done using different
interactions for nucleon, $K^-$ and $\bar K^0$ phases. With each parameter
set, as we increase $U_K$, the softening effect of anikaons on the EoS is more and hence the maximum mass decreases
rapidly. When we are not considering the antikaon phase, the maximum mass   corresponding to G1 and FSU2.1 parameters are same, but inclusion of antikaons  changes the results completely. From Fig.~\ref{fig:massradius}
it is also evident that at a particular $U_K$, change in the maximum mass
($\Delta M$) corresponding
to NL3 is more and that for FSU2.1 is less and $\bar K^0$ does not play any role in case of
FSU2.1 parameter set. In the presence of $K^-$, factors governing the change in maximum mass are discussed in our previous work \cite{neha} and the presence
of $\bar K^0$ depends mostly on the onset and presence of $K^-$.
From Fig.~\ref{fig:massradius}(b), one can observe that the radius corresponding to maximum mass is lower for a softer EoS and this can change in the presence of antikaons. Presence of antikaons soften the EoS but they can increase or decrease the central baryon density ($\rho_c$) and hence the radius ($R$). We can summarise the effect of $U_K$,  on the mass and radius as, 
$\Delta M\propto U_K$ and $R\propto 1/\rho_c$.

\section{Summary} We extend different versions of relativistic mean field
(RMF)   Langragians (RMF, E-RMF
and FSU2.1), with the inclusion of  antikaons. We analyze the role of antikaons,
in conjunction with additional (higher-order) couplings in the recent RMF
models,  on the neutron star properties.
The additional couplings  soften the EoS for symmetric, pure neutron, and neutron
star matters and also delay the onset of both the antikaons in the neutron star.
We conclude that additional couplings in RMF, which play a significant role at higher densities, are also important where antikaons dominate the behavior of equation of state. 

\section*{Acknowledgments}
Neha Gupta acknowledges CSIR and CICS, Government of India, for providing financial
support to present this work in the conference NN2012, held at San Antiono, Texas, USA. 

\providecommand{\newblock}{}


\begin{thebibliography}{10}
\expandafter\ifx\csname url\endcsname\relax
  \def\url#1{{\tt #1}}\fi
\expandafter\ifx\csname urlprefix\endcsname\relax\def\urlprefix{URL }\fi
\providecommand{\eprint}[2][]{\url{#2}}

\bibitem{lattimer:426}
Lattimer J~M and Prakash M 2001 {\em Astrophys. J.\/} {\bf 550} 426

\bibitem{gle}
Glendenning N~K 2007 {\em Compact Stars\/} 2nd ed (Springer-Verlag, New York)

\bibitem{expt_m}
Champion D~J {\em et~al.\/} 2008 {\em Science\/} {\bf 320} 1309

\bibitem{demorest}
Demorest P~B, Pennucci T, Ransom S~M, Roberts M~S~E and Hessels J~W~T 2010 {\em
  Nature\/} {\bf 467} 1081

\bibitem{FSU_para}
Todd-Rutel B~G and Piekarewicz J 2005 {\em Phys. Rev. Lett.\/} {\bf 95} 122501

\bibitem{ermf}
Arumugam P, Sharma B~K, Sahu P~K, Patra S~K, Sil T, Centelles M and Vi\~nas X
  2004 {\em Phys. Lett. B\/} {\bf 601} 51

\bibitem{furnstahl}
Furnstahl R~J, Serot B~D and Tang H~B 1997 {\em Nucl. Phys. A\/} {\bf 615} 441

\bibitem{neha}
Gupta N and Arumugam P 2012 {\em Phys. Rev. C\/} {\bf 85} 015804

\bibitem{Banik035802}
Banik S and Bandyopadhyay D 2001 {\em Phys. Rev. C\/} {\bf 63} 035802

\bibitem{Pal553}
Pal S, Bandyopadhyay D and Greiner W 2000 {\em Nuclear Physics A\/} {\bf 674}
  553

\bibitem{Kap}
Kaplan D~B and Nelson A~E 1986 {\em Phys. Lett. B\/} {\bf 175} 57 -- 63

\bibitem{Ellis199511}
Ellis P~J, Knorren R and Prakash M 1995 {\em Physics Letters B\/} {\bf 349} 11
  -- 17

\bibitem{Brown:355}
Brown G, Kubodera K, Rho M and Thorsson V 1992 {\em Physics Letters B\/} {\bf
  291} 355 -- 362

\bibitem{Thorsson1994693}
Thorsson V, Prakash M and Lattimer J~M 1994 {\em Nuclear Physics A\/} {\bf 572}
  693 -- 731

\bibitem{banik:055805}
Banik S and Bandyopadhyay D 2001 {\em Phys. Rev. C.\/} {\bf 64} 055805

\bibitem{glen_kaon}
Glendenning N~K and Schaffner-Bielich J 1999 {\em Phys. Rev. C\/} {\bf 60}
  025803

\bibitem{nl3_para}
Lalazissis G~A, K\"onig J and Ring P 1997 {\em Phys. Rev. C\/} {\bf 55} 540

\bibitem{Shen:065808}
Shen G, Horowitz C~J and O'Connor E 2011 {\em Phys. Rev. C\/} {\bf 83} 065808

\bibitem{expt_data}
Danielewicz P, Lacey R and Lynch W~G 2002 {\em Science\/} {\bf 298} 1592

\bibitem{expt_m5}
\"Ozel F, Baym G and G\"uver T 2010 {\em Phys. Rev. D\/} {\bf 82} 101301

\bibitem{Fattoyev:055803}
Fattoyev F~J, Horowitz C~J, Piekarewicz J and Shen G 2010 {\em Phys. Rev. C\/}
  {\bf 82} 055803

\bibitem{kappa_4}
Gueorguiev V~G, Ormand W~E, Johnson C~W and Draayer J~P 2002 {\em Phys. Rev.
  C\/} {\bf 65} 024314

\bibitem{tov1}
Tolman R~C 1939 {\em Phys. Rev.\/} {\bf 55} 364

\bibitem{tov}
Oppenheimer J~R and Volkoff G~M 1939 {\em Phys. Rev.\/} {\bf 55} 374

\end{thebibliography}
\end{document}